\documentclass[aps,pra,twocolumn,groupedaddress,showpacs,showkeys]{revtex4}

\usepackage{graphicx}

\begin{document}

\newcommand{\be}{\begin{equation}}
\newcommand{\ee}{\end{equation}}

\title{Isolelectronic apparatus to probe the thermal Casimir force.}

\date{\today}

\author{Giuseppe Bimonte}
\affiliation{Dipartimento di 
Fisica, Universit{\`a} di Napoli Federico II, Complesso Universitario
MSA, Via Cintia, I-80126 Napoli, Italy}
\affiliation{INFN Sezione di
Napoli, I-80126 Napoli, Italy }

\begin{abstract}

Isoelectronic differential force measurements provide a unique opportunity to probe controversial features of the thermal Casimir effect, that are still  much debated in the current literature. Isolectronic setups offer two major advantages over conventional Casimir setups.   On one hand  they are  immune from  electrostatic forces caused by potential patches on the plates surfaces, that plague present Casimir experiments especially for separations in the micron range. On the other hand they can strongly enhance the discrepancy  between alternative theoretical models that have been proposed to estimate the thermal Casimir force for metallic and magnetic surfaces. Thanks to these two features, isoelectronic differential experiments  should allow to establish conclusively which among these models correctly describes the thermal Casimir force.

\end{abstract}

\maketitle

\section{Introduction}

Over sixty years ago \cite{Casimir48} the dutch physicist Hendrik Casimir, building on the basic principles of Quantum Electrodynamics, predicted that two discharged perfectly conducting plane-parallel surfaces in vacuum attract each other with a (unit-area) force of magnitude 
\be
F_C=\frac{\pi^2 \hbar c}{240\,a^4} \;.
\ee  
This  force orginates from zero-point quantum fluctuations of the electromagnetic (em) field that according to Quantum Theory fill empty space, or more precisely from modifications of the spectrum of these fluctuations caused by the presence of polarizable (but otherwise neutral) material surfaces.  In his pioneering paper, Casimir considered two perfectly conducting plates at zero temperature. The theory of the Casimir effect for real material surfaces was developed a few years later by Lifshitz \cite{lifs}, by extending to dispersion forces Rytov's theory of electromagnetic fluctuations \cite{rytov}.
For  a  general overview on the Casimir effect  see \cite{book1,book2,parse,revexp}.

During the fifty years following Casimir's seminal paper, a few experiments were performed to observe the Casimir force which, apart from providing a qualitative confirmation of the effect, had the important merit of identifying the experimental problems that had to be addressed for a successful observation of the tiny Casimir force.    
The modern era of the Casimir effect started in 1997 with a landmark torsion-balance experiment by S. Lamoreaux \cite{lamor1}, soon followed by the AFM experiment of Mohideen and Roy       \cite{umar}. These experiments opened the era of precision Casimir experiments, by which it became possible for the first time to explore a number of features of the Casimir effect relating to material properties of the involved surfaces, that could be predicted on the basis of Lifshitz theory.  Several other experiments followed in rapid succession, which utilized both metallic surfaces  \cite{gianni,decca4,decca5,decca6,chang2}, as well as surfaces made of  diverse materials like   semiconductors \cite{semic},   conductive oxides \cite{ito}, ferromagnetic metals \cite{bani1, bani2}, and surfaces immersed in  liquids \cite{liq}. Superconducting Casimir devices have been studied as well \cite{ala1,ala2,super1,superc}.  Much excitement was spurred  by the possible exploitation of the Casimir force in the actuation of  micro and nano machined devices \cite{capasso}, which stimulated investigations of the Casimir effect for microstructured surfaces \cite{chen,chan,bao,chiu,bani,deccanat}. For  a review   of these and many other experiments see \cite{book1,book2,revexp,capassorev}.

The specific subject of this paper is the effect on the Casimir force of the finite temperature $T$ of the plates, a  problem that has attracted a good deal of interest in recent years. In his seminal paper, Lifshitz showed that thermal fluctuations of the em field provide an extra contribution to the Casimir force, called {\it thermal} Casimir force.  
Surprisingly,  estimating the magnitude of the thermal force for conducting plates turned into an intriguing puzzle. In essence the puzzle is about the role played by relaxation properties of conduction electrons in Lifshitz theory. It turns out that Lifshitz formula predicts significantly different magnitudes for the thermal force depending on whether   the optical data of the conductor are extrapolated towards zero frequency on the basis of the Drude model (which does take dissipation into account) or instead by the dissipationless plasma model of IR optics.     
In addition to predicting different magnitudes for the thermal force, it has been shown that the two prescriptions have important thermodynamic consequences: while the Drude prescription leads to a violation of Nernst heat theorem (in the idealized case of two conducting plates with a perfect crystal structure) \cite{bezerra,bezerra2}, the plasma prescription violates the Bohr-van Leeuwen theorem of classical statistical physics \cite{Martin,bimo2}. 

The experimental situation is contradictory. Several small distance experiments \cite{bani2,decca4,decca5,decca6,chang2}, probing separations below one micron,  appear to be in agreement with the plasma model, and to rule out the larger thermal force predicted by the Drude model. These experiments provide the most precise measurements of the Casimir force to date, with errors in the percent range.  One has to bear in mind however that  for separations $a$  smaller than the thermal length  $\lambda_T=
\hbar c/(2 \pi k_B T)$ (for $T=300$ K $\lambda_{\rm 300 K}=1.2\; \mu$m)  the thermal force only represents a small correction to the zero-point force, and therefore the theoretical interpretation of these experiments is very delicate. For a review of these experiments
see \cite{critiz}.  

In principle,  observing  the thermal Casimir force should be easier for   separations $a \gtrsim \lambda_T$ because 
for these large separations the thermal force is dominant. Discriminating between the Drude and the plasma models should be easier as well, because for large distances  the two prescriptions  predict markedly different magnitudes for the Casimir force. For example, for two plane-parallel conducting surfaces  the Drude model predicts for $a \gg \lambda_T$  a Casimir pressure of magnitude  $\zeta(3) k_B T /8 \pi a^3$, while the plasma model predicts a magnitude twice as large. Unfortunately observing the thermal force for separations in the micron region is very difficult too, not only because the Casimir force quickly gets very small as the separation increases, but also because of unavoidable stray electrostatic forces that mask the Casimir force.   These stray forces, that cannot be eliminated by applying a bias potential, originate from patches of varying potential on the surfaces caused by spatial changes of crystalline structure and/or by adsorbed impurities. Stray electrostatic forces over 100 times  stronger than the   thermal Casimir force  were indeed observed  in an experiment with Al surfaces\cite{antonini} in the range from 3.5 to 5 $\mu$m.  Large electrostatic forces were reported as well in a recent experiment by the Yale group \cite{lamorth}, which claims to have observed the thermal force between a large sphere and a plate both covered with gold, in the wide range of separations form 0.7 to 7.3 $\mu$m. The results have been interpreted by the authors as being in accordance with the  Drude prescription. This experiment has been criticized  \cite{critiz}, because the thermal Casimir force was obtained only after subtracting from the total measured force the much larger electrostatic force.
The subtraction was perfomed by making a fit of the total observed force, based on a two-parameter model of the electrostatic force,   and not by a direct and independent measurement, as it would have been desirable. The problem of patch potentials  is regarded as a major obstacle for present Casimir experiments,  and dedicated techniques based on Kelvin probe force microscopy are being developed to achieve a direct observation of the patches with the necessary spatial resolution \cite{deccapatch}.
       
The contradictory results of recent experiments call for new experiments specifically designed to probe the thermal Casimir force. Recently the author proposed two setups \cite{hide, hide2} that should allow for an unambiguous observation of the thermal Casimir force. 
Both setups are based on isoelectronic differential force measurements, an idea pioneered by the  IUPUI group in searches for non-newtonian gravitational forces in the sub-micron region \cite{deccaiso,decca7}. The unique virtue of this approach is that it is immune from the problem of patch potentials that plague conventional Casimir experiments, especially for large separations. A  thorough analysis of the limitations on the sensitivity of  isolelectronic experiments resulting from random spatial variations of patch forces  has been recently carried out \cite{speake}, confirming the high suppression of patch forces in isoelectronic setups. In \cite{hide, hide2} it was proved that the isolelectronic technique provides a powerful  tool to observe the elusive thermal Casimir force. In this paper we further develop the findings of \cite{hide, hide2}, and we demonstrate that by  suitably choosing the materials of the samples   it is possible to strongly enhance the discrepancy between the Drude and the plasma models,  both for non-magnetic and for magnetic metals. This makes one confident that by this technique  it  should be possible to establish conclusively  which among these models  correctly describes the thermal Casimir force. In particular, it should be possible to  clarify  if and to what extent  the large magnetic permeability of ferromagnetic materials influences the Casimir force, a problem that has been investigated recently by the Riverside group \cite{bani1,bani2}.   For ferromagnetic metals the isolectronic scheme  is especially effective, because with such a setup the relative difference between alternative models of the thermal Casimir force can be    as large as one thousandt \cite{hide}. Preliminary  results of an ongoing experiment at IUPUI  based on the scheme of \cite{hide},  already prove conclusively that in the case of Ni the observed signal is three orders of magnitude smaller than the theoretical prediction  based on the  Drude model. In this paper we show how isoelectronic setups  can be designed to further  investigate whether magnetic properties affect at all the Casimir force.  

The plan of the paper is as follows.  In Sec. II we review the Drude and the plasma prescriptions for computing the thermal Casimir force. In Sec. III we present the general structure of isoelectronic setups  for Casimir experiments, while in Sec. III-A and III-B we describe in detail  setups specifically designed to investigate the thermal Casimir effect for non-magnetic  and for magnetic metals, respectively. Finally Sec. IV presents our conclusions.

\section{Drude and plasma prescriptions}

According to Lifshitz theory \cite{lifs}, the Casimir free-energy  ${\cal F}$ (per unit area) between two dieletric  plane-parallel slabs $S_{j}$, $j=1,2$  at distance $a$ in vacuum   is given by the formula: 
$$
{\cal F}(T,a)=\frac{k_B T}{2 \pi}\sum_{l=0}^{\infty}\left(1-\frac{1}{2}\delta_{l0}\right)\int_0^{\infty} d k_{\perp} k_{\perp}  
$$
\be
\times \; \sum_{\alpha={\rm TE,TM}} \log \left[1- {e^{-2 a q_l}}{R^{(1)}_{\alpha}({\rm i} \xi_l,k_{\perp})\;R^{(2)}_{\alpha}({\rm i} \xi_l,k_{\perp})} \right]\;,\label{lifs}
\ee
where $k_B$ is Boltzmann constant, $\xi_l=2 \pi l k_B T/\hbar$ are the (imaginary) Matsubara frequencies, $k_{\perp}$ is the modulus of the in-plane wave-vector, $q_l=\sqrt{\xi_l^2/c^2+k_{\perp}^2}$, and $R^{(j)}_{\alpha}({\rm i} \xi_l,k_{\perp})$ is the familiar Fresnel reflection coefficient of slab $j$ for polarization $\alpha$: 
\be
R^{(j)}_{\rm TE}=\frac{q_l-  \,k_l^{(j)}}{ q_l+ \,k_l^{(j)}}\;,\label{freTE}
\ee
\be
R^{(j)}_{\rm TM}=\frac{\epsilon_{j} ({\rm i} \xi_l) \,q_l- \,k_l^{(j)}}{\epsilon_{j}({\rm i} \xi_l) \,q_l+  \,k_l^{(j)}}\;,\label{freTM}
\ee
where
$ k_l^{(j)}=\sqrt{\epsilon_j({\rm i} \xi_l)  \xi_l^2/c^2+k_{\perp}^2}$,  and
 $\epsilon_j$ is the (dynamic) electric   permittivity of slab $j$.   According to Lifshitz formula, to compute the Casimir energy one needs to know the permittivty $\epsilon({\rm i} \xi)$ of the involved materials along the imaginary frequency axis. This quantity cannot of course be measured directly,  but it can be computed using dispersion relations  on the basis of optical data referring to real frequencies $\omega$. For the case of insulators or ohmic conductors, the standard formula is provided by the Kramers-Kronig dispersion relation which expresses  $\epsilon({\rm i} \xi)$ in terms of the imaginary part $\epsilon''(\omega)$ of the  permittivity:
 \be
\epsilon( {\rm i} \xi)=1+\frac{2}{\pi}\int_0^{\infty} d \omega
\frac{\omega\,
\epsilon''(\omega)}{\omega^2+\xi^2}\;.\label{disp}\ee
We see from Eq. (\ref{disp}) that in order to
evaluate $\epsilon(i \xi)$ at any   imaginary frequency $\xi$  it
is in principle necessary to know $\epsilon''(\omega)$ at all
frequencies $\omega$. Unfortunately such a complete knowledge of
$\epsilon''(\omega)$ is never possible, because optical data are
always   restricted to some finite frequency range $\omega_{\rm
min} < \omega < \omega_{\rm max}$, starting from a non-zero
minimum frequency $\omega_{\rm min}>0$. In practice, there is no  real difficulty on the high frequency
side, because fall-off properties of $\epsilon''(\omega)$ ensure
that for the relevant $\xi$'s real frequencies $\omega$
larger than a few tens of eV/$\hbar$ give already a negligible
contribution to the integral on the r.h.s. of Eq. (\ref{disp}). On the low-frequency side, however, one faces a  problem when metals are considered.
Since the
imaginary part of the permittivity of ohmic conductors diverges like $1/\omega$ for small frequencies,  the integral in Eq.
(\ref{disp}) receives a large contribution from low
frequencies for which no optical data are available.    Since
truncation of the integral to the frequency $\omega_{\rm min}$
would result in a large error,
one is forced to extrapolate the dielectric function
$\epsilon''(\omega)$ to frequencies $\omega < \omega_{\rm min}$,
where optical data are not available, to evaluate the integral for
$\omega<\omega_{\rm min}$.  As a rule, the extrapolation is done
using the simple Drude model \be \epsilon_{\rm
Dr}(\omega)=1-\frac{\omega_p^2}{\omega(\omega+{\rm  i}
\gamma)}\;,\label{drude}\ee where $\omega_p$ is the plasma
frequency, and $\gamma$ is the relaxation frequency. According to the Drude prescription, $\epsilon({\rm i} \xi)$ is then estimated by the formula
\be
\epsilon( {\rm i} \xi)=1+\frac{2}{\pi}\int_0^{\omega_{\rm min}} d \omega
\frac{\omega\,
\epsilon_{\rm Dr}''(\omega)}{\omega^2+\xi^2} +\frac{2}{\pi}\int_{\omega_{\rm min}}^{\infty} d \omega
\frac{\omega\,
\epsilon''(\omega)}{\omega^2+\xi^2}\;,\label{disp2}\ee
where $\epsilon''_{\rm Dr}(\omega)$ is the imaginary part of the Drude permittivity Eq. (\ref{drude}).
The error introduced by this extrapolation   in the estimate of $\epsilon({\rm  i} \xi)$, and thereof of the Casimir force,   has been a subject of intense discussion \cite{Piro,sveto}. The problem can be partly relieved by using weighted dispersion relations \cite{generKK, generKK2}, which sensibly reduce the contribution of the extrapolation.    

\begin{figure}
\includegraphics{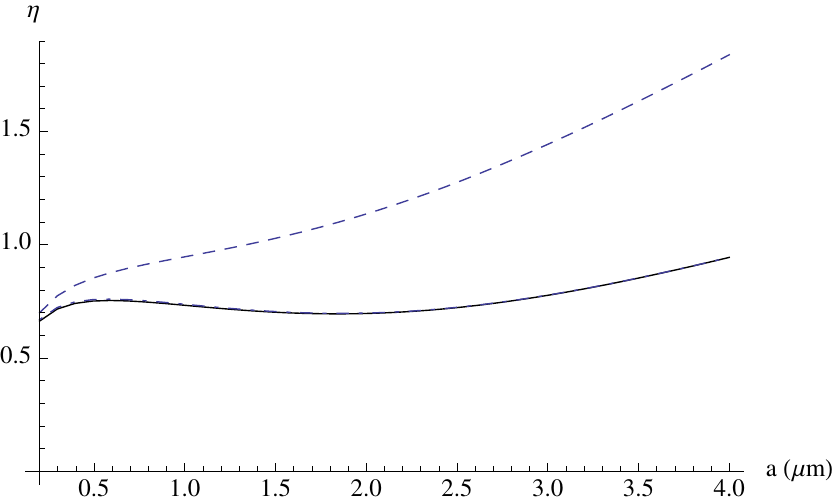}
\caption{\label{eta}  Reduction factor $\eta={\cal F}/{E_{\rm id}}$ for two gold plates at room temperature, versus plates separation (in micron). The solid line is computed using the Drude prescription, while the dashed line is for the plasma prescription. The dot-dashed line corresponds to the plasma prescription,
after subtraction of the contribution of the TE $l=0$ mode. }
\end{figure}

Surprisingly the estimate of   $\epsilon({\rm  i} \xi)$ obtained by the above procedure, based on a Drude extrapolation of the optical data,  when plugged into Lifshitz formula results in a prediction of the Casimir force that  appears to be inconsistent with several recent experiments  \cite{bani2,decca4,decca5,decca6,chang2}. It has been claimed that  a prediction of the Casimir force which is consistent with the data  can be obtained if   $\epsilon({\rm  i} \xi)$ is computed by a different procedure, in which relaxation properties of conduction electrons are neglected altogether. According to this so-called plasma prescription $\epsilon({\rm  i} \xi)$ is computed by the formula:
 \be
\epsilon( {\rm i} \xi)=1+\frac{\omega_p^2}{\xi^2}+\frac{2}{\pi}\int_{\omega_{\rm min}}^{\infty} d \omega
\frac{\omega\,(
\epsilon''(\omega)-\epsilon''_{\rm Dr}(\omega))}{\omega^2+\xi^2}\;.\label{disp3}\ee
 The main difference between the two prescriptions is in the power of divergence of $\epsilon({\rm i} \xi)$ for vanishing $\xi$. While with the Drude prescription $\epsilon({\rm i} \xi)$ has a simple pole, characteristic of ohmic conductors:
\be
\epsilon({\rm i} \xi) = \frac{\omega_p^2}{\gamma \xi}+ O(\xi^0)\;\;\;\;({\rm Drude\; prescription}),
\ee    
the plasma prescription leads to a double pole, similarly to superconductors \cite{super2}:
\be
\epsilon({\rm i} \xi) = \frac{\omega_p^2}{\xi^2}+ O(\xi^0)\;\;\;\;({\rm plasma\; prescription})\;.
\ee    
When the two prescriptions are used to compute the Casimir free energy between two metallic plates for room temperature, one gets almost coinciding values for the contributions of the non-vanishing Matsubara terms (i.e. the terms with $l \neq 0$ in Eq. (\ref{lifs})). The $l=0$ term with TM polarization is of course the same, because for both prescriptions the zero-frequency reflection coefficient for TM modes is one. The major difference is seen in the $l=0$ TE mode,  since this mode vanishes according to the Drude prescription while it does not according to the plasma prescription, as a result of the different singular behaviors of $\epsilon({\rm i} \xi) $ for vanishing $\xi$.  To give the reader a feeling of the magnitude of the  difference among the Drude and plasma predictions, in Fig. (\ref{eta}) we plot the  corresponding reduction factors $\eta={\cal F}/{E_{\rm id}}$ for two gold plates, where $E_{\rm id}=-\pi^2 \hbar c/(720 a^3)$ is the $T=0$ energy (per unit area) between two perfectly conducting parallel plates.   The solid line in Fig. (\ref{eta})  is computed using the Drude prescription, while the dashed line is for the plasma prescription. The dot-dashed line corresponds to the plasma prescription,
after subtraction of the contribution of the TE $l=0$ mode.  The solid line is barely distinguishable from the dot-dashed line, which shows as anticipated earlier that the difference between the Drude and the plasma prescriptions arises solely from the thermal TE $l=0$ mode.   
The plot shows also that the largest difference between the two prescriptions is found for separations larger than one micron, where the thermal force becomes dominant. Compared to the plasma model, the Drude prescription   predicts an extra repulsive thermal force which, was has been shown to originate from the magnetic interaction among thermal Focault currents existing in metallic plates an non-zero temperature \cite{john,carsten}.

The above picture of the Drude vs. plasma comparison, described  here for the case of two non-magnetic dielectric slabs, remains qualitatively true in the more general situations considered later in this work, when we move on to layered slabs and/or magnetic materials: in all cases the  different magnitudes of the thermal force predicted by the Drude and the plasma prescriptions   originate from the TE $l=0$ mode, because the two prescriptions  entail markedtly different values for the reflection coefficient of the TE mode for vanishing frequency.       

\section{Isoelectronic Casimir setups}

\begin{figure}
\includegraphics{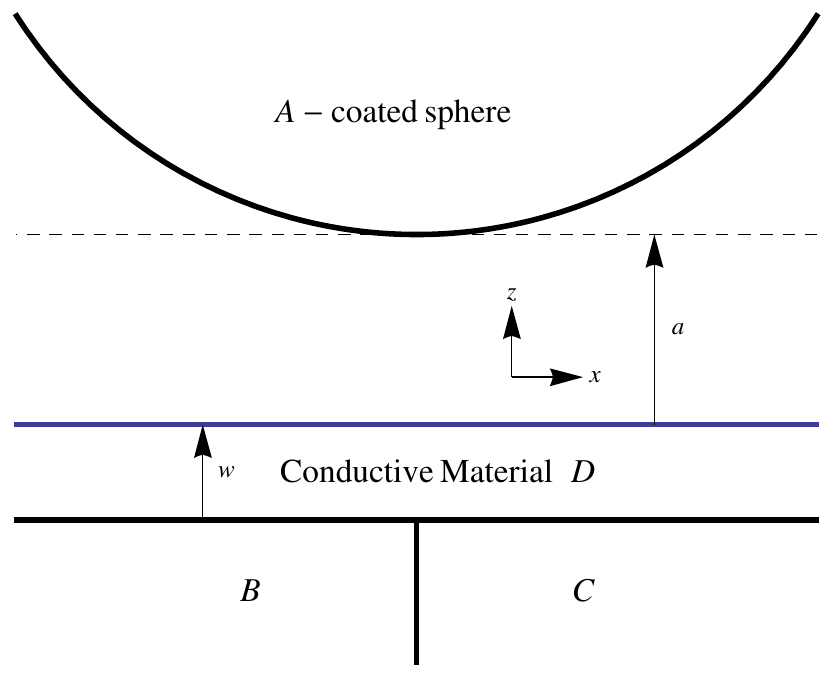}
\caption{\label{setup}   The isoelectronic setup consists of a  sphere coated with material A   above a  planar slab divided in two regions,  respectively made of materials B and C. Both the B and C regions are covered  with a semi-transparent plane-parallel conducting over-layer of uniform thickness $w$ made of material D. }
\end{figure}

 
As we discussed earlier, patch potentials constitute a major obtacle towards the observation of the thermal Casimir force. 
In recent searches of non-newtonian gravity \cite{deccaiso,decca7} the IUPUI group developed an elegant experimental technique based on isoelectronic force-difference measurements, which is by design immune from the problem of potential patches.  The structure of the isoelectronic setups we consider in this work is schematically illustrated in Fig. \ref{setup}. It  consists of a sphere  of radius $R$ covered by a coating of material A  and a planar slab divided in two regions,  made of  two different materials B and C, respectively. 
The key feature of the isoelectronic apparatus is the  {\it conductive} over-layer of thickness $w$ made of material D, covering both the B and the C regions. 
For any fixed sphere-plate separation $a$, we consider measuring  the {\it difference} 
\be\Delta F(a)=F_{\rm B}(a)-F_{\rm C}(a)\ee among  the values $F_{\rm B}$ and $F_{\rm C}$  of  the (normal) Casimir force that obtain when the tip of the sphere is respectively above a point $q$   deep in the B region, and a point $p$   deep in the C region. The great advantage of this differential mesurement over an absolute force measurement is that the  detrimental (mean) electrostatic force caused by patches on the exposed surfaces of the plates is automatically subtracted out from $\Delta F$, provided of course that the surface of the over-layer  has identical patch structure above the two regions of the plate.  The limitations on the sensitivity of  isolelectronic experiments resulting from random spatial variations of patch forces  have  been recently studied in detail \cite{speake}, confirming the high suppression of patch forces in isoelectronic setups. 
An important difference between our setups and those  used by the IUPUI group  should be stressed, however. In order to observe small differences between the gravitational interaction of the sphere with the  B and C sectors of the plate, the conductive overlayer of \cite{deccaiso,decca7} was designed to be opaque, to screen out altogether the otherwise dominant electrostatic and Casimir forces.  The thickness ($w=150$ nm) of the Au overlayers used in those experiments was therefore chosen to be {\it  larger} than the plasma length of Au  ($\lambda_p$=130 nm). In our setup, instead, we do want to observe the differential thermal Casimir interaction of the sphere with the B and C regions of the plate. The conducting over-layer should then be designed such as to screen out the unwanted electrostatic component of the force, but at the same time it should impede as little as possible the passage of thermal photons that are responsible of the thermal Casimir interaction with the underlying materials. To do that, our overlayer has to be semi-transparent to infra-red radiation, and therefore its thickness has to be {\it smaller} than the skin depth of thermal photons for material D. 

We  assume, as it is usually the case in Casimir experiments, that the sphere radius $R$ is much larger than the  separation $a$. Under this condition, it is possible to   estimate the Casimir force by the Proximity Force Approximation (PFA).  The PFA has been widely used to interpret most  Casimir experiments \cite{book2}  (see  \cite{kruger} for more applications of the Proximity Approximation). 
Recently, it has been shown that the PFA represents  the leading term in a gradient expansion of the Casimir force, in powers of the slopes of the bounding surfaces \cite{fosco2,bimonte3,bimonte4}. Very recently the gradient expansion has been used to compute curvature corrections to the Casimir-Polder potential between a small polarizable body and a curved surface \cite{bimonteCP}.  The gradient expansion shows that the PFA is asymptotically exact in the zero-curvature limit, and with its help it is now possible  to estimate the error introduced by the PFA.  
Besides $R \gg a$, two further conditions are required to ensure the validity of the PFA in our setup.  To be definite, we let $(x,y,z)$ be cartesian coordinates  such that $(x,y)$   span the exposed surface    of the conductive overlayer, placed at $z=0$, while the sphere tip is at $z=a$. We imagine that the $x<0$ region of the slab is made of material B, while its $x>0$ region is made of material C.  
To ensure that we can neglect the effect of the sharp boundary between the B and the C regions, we 
assume that the horizontal distances of the points $p$ and $q$  from the B-C boundary are both much larger than the typical interaction radius $\rho=\sqrt{a R}$ of the circular region of the plate that contributes significantly to the Casimir interaction: $s \gg \rho$.  The force $F_{\rm B}$ ($F_{\rm C}$) is then undistinguishable from the force ${\tilde F}_{\rm B}$ (${\tilde F}_{\rm C}$) between the  sphere and a plane-parallel two-layers slab consisting of  a  layer of thickness $w$ of material D on top of an infinite dielectric  planar slab made of material B (C).  
Under these conditions, we  have for $\Delta F$:
$$
\Delta F \simeq {\tilde F}_{\rm B}- {\tilde F}_{\rm C} \simeq {F}_{\rm B}^{(\rm PFA)}-{F}_{\rm C}^{(\rm PFA)}
$$
\be
=2 \pi R ({ {\cal F}}_{\rm B}- { {\cal F}}_{\rm C})\;,\label{forpfa}
\ee 
where ${ { F}}_{\rm B}^{(\rm PFA)}$ and ${ { F}}_{\rm C}^{(\rm PFA)}$ denote the PFA expressions for the Casimir force between the sphere and a two-layer slab consisting of   the conductive layer of material D on top of either a B or a C substrate, respectively, while  ${ {\cal F}}_{\rm B} $ and ${\tilde {\cal F}}_{\rm C} $ denote the unit-area Casimir free energies  for the corresponding plane-parallel systems in which the sphere is replaced by a  planar slab of material A. In the last passage of Eq. (\ref{forpfa}) we used the well-known PFA formula for the Casimir free energy of a sphere-plate system $
F^{(\rm PFA)}_{\rm sp-pl}(a) = 2 \pi R {\cal F}(a) $.  For later applications, we consider the possibility that some among the materials A, B and C are magneto-dielectric.   In its original version Lifshitz theory was formulated for
  planar dielectrics ($\mu = 1$) fully described by the respective frequency dependent
(complex) dynamical permittivity $\epsilon(\omega)$. The theory was later generalized to deal with layered magneto-dielectric plates in \cite{richmond,tomas}.
It turns out that the Casimir free-energy  ${\cal F}$ (per unit area) between two magnetodieletric possibly layered plane-parallel   slabs $S_{j}$, $j=1,2$ at distance $a$ in vacuum   is still represented by Eq. (\ref{lifs}), provided that the reflection coefficients are now understood to be those for the possibly layered slabs.
The Casimir free energy ${\cal F}_{\rm B}$ can be  obtained from  Eq. (\ref{lifs}) by substituting  $R^{(1)}_{\alpha}$ by the Fresnel reflection coefficient $r_{\alpha}^{(0{\rm A})}$   (given in Eqs.(\ref{freTE}) and (\ref{freTM}) below, with $a=0$, $b=$A),  and $R^{(2)}_{\alpha}$ by the reflection coefficient  $R_{\alpha}^{(0{\rm DB})}$ of a two-layer planar slab consisting of a layer of thickness $w$ of material D on a dielectric B substrate.  The latter reflection coefficient has the expression:
\be  
R_{\alpha}^{(0{\rm D B})}({\rm i} \xi_l,k_{\perp};w)=\frac{r_{\alpha}^{(0{\rm D})}+e^{-2\,w\, k_l^{({\rm D})}}\,r_{\alpha}^{({\rm D B})}}{1+e^{-2\,w\, k_l^{({\rm D})}}\,r_{\alpha}^{(0{\rm D})}\,r_{\alpha}^{({\rm D B})}}\;.
\ee   
\be
r^{(ab)}_{\rm TE}=\frac{\mu_{b}({\rm i} \xi_l) \,k_l^{(a)}-\mu_{a}({\rm i} \xi_l) \,k_l^{(b)}}{\mu_{b}({\rm i} \xi_l) \,k_l^{(a)}+\mu_{a}({\rm i} \xi_l) \,k_l^{(b)}}\;,\label{freTE}
\ee
\be
r^{(ab)}_{\rm TM}=\frac{\epsilon_{b}({\rm i} \xi_l) \,k_l^{(a)}-\epsilon_{a}({\rm i} \xi_l) \,k_l^{(b)}}{\epsilon_{b}({\rm i} \xi_l) \,k_l^{(a)}+\epsilon_{a}({\rm i} \xi_l) \,k_l^{(b)}}\;,\label{freTM}
\ee
where
$ k_l^{(a)}=\sqrt{\epsilon_a({\rm i} \xi_l) \mu_a({\rm i} \xi_l) \xi_l^2/c^2+k_{\perp}^2}\;$, 
 $\epsilon_a$ and $\mu_a$ denote the electric and magnetic permittivities of medium $a$, and we define $\epsilon_0=\mu_0=1$.
The Casimir free energy ${\cal F}_{\rm C}$ for a planar A-${\rm D}$-C system
can be obtained by replacing material B by material C in the above formulae.

\subsection{Non-magnetic metals}

The first setup we consider is designed to discriminate among the Drude and the plasma prescriptions for non-magnetic conductors. For this purpose  we found that a convenient choice of materials is the following: we take Au for the coating of the sphere and for the C region of the plate, and   high-resistivity (dielectric) Si for the region B.  The choice of the material for the overlayer is critical. As we explained earlier, the overlayer needs to be semitransparent to thermal photons.  This requirement led us to consider   B-doped {\it low-resistivity} Si,   because its large plasma length ($\lambda_p=2.7 \mu$m)  ensures a good degree of transparency also for relatively large thicknesses $w$ (we take $w=100$ nm).  B-doped Si has been used successfully in Casimir experiments (see the second of Refs. \cite{semic}). Below, we shall use the symbol ${\rm Si_c}$ to denote conductive silicon, while the symbol Si shall denote dielectric silicon. 

The values $\epsilon_{\rm a}({\rm i} \xi_l)$, (a=Au, Si)  of the permittivities of Au and dielectric Si were computed by  the procedure expalined in Sec. II,
on the basis of the tabulated optical data quoted in \cite{palik}. When using the Drude prescription, the permittivity $\epsilon_{\rm Au}({\rm i} \xi)$ of Au  was computed according to  Eq. (\ref{disp2}).  The optical data for Au were extrapolated towards low frequencies by the Drude model, with Drude parameters  $\omega_{\rm Au}=8.9 \,{\rm eV}/\hbar$,  $\gamma_{\rm Au}=0.035\, {\rm eV}/\hbar$. For the permittivity of conductive Si  we used the formula
\be \epsilon_{\rm Si_c}({\rm i} \xi)=\epsilon_{\rm Si}({\rm i} \xi)+\frac{\omega_{{\rm Si_c}}^2}{\xi(\xi + \gamma_{\rm Si_c})}\;, \label{condsil}\ee with plasma frequency $\omega_{{\rm Si_c}}=0.46\,{\rm eV}/\hbar$ 
and relaxation frequency $ \gamma_{\rm Si_c}=0.1\,{\rm eV}/\hbar$  (see Ref. \cite{book2}, pag. 588). When we considered the plasma prescription, we made use of Eq. (\ref{disp3}) to compute the imaginary-frequency permittivity of Au. In the case of ${\rm Si_c}$ the plasma prescription was implemented simply by setting to zero $\gamma_{\rm Si_c}$ in Eq. (\ref{condsil}) . 
 
\begin{figure}
\includegraphics{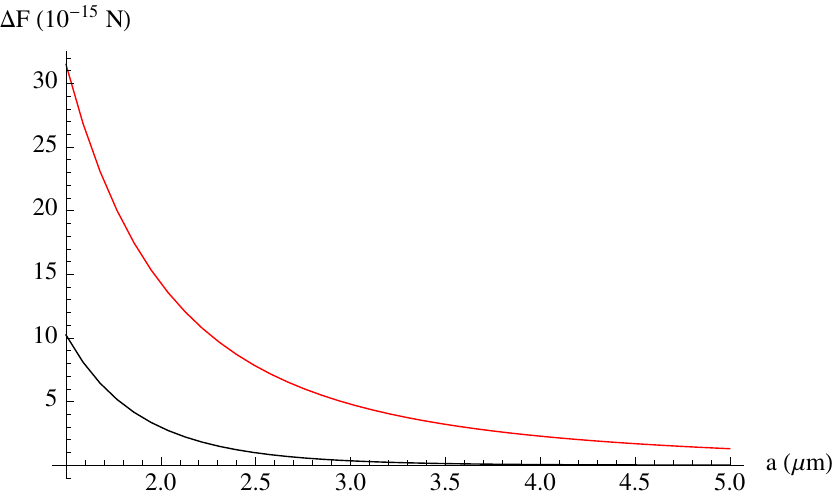}
\caption{\label{fig2}  Force difference $\Delta F$ (in fN) versus separation $a$ (in microns) for a Au sphere of radius $R=150\;\mu$m  in front of a Si-Au plate, covered with 100 nm of low-resistivity Si.  The red and grey lines correspond to  the plasma and Drude prescriptions, respectively. }
\end{figure}

In Fig. \ref{fig2} we plot   $\Delta F $   (in fN) versus separation $a$ (in $\mu$m) for a sphere of radius $R=150\; \mu$m and for a thickness $w=100$ nm of the ${\rm Si_c}$ over-layer. The red and the grey curves correspond the plasma and  Drude prescriptions, respectively. The plasma and the Drude models predict widely different magnitudes for $\Delta F$. For example, for $a=3$ $\mu$m $\Delta F_{\rm plasma}$ is fourteen times larger than $\Delta F_{\rm Drude}$, while for $a=4$ $\mu$m they differ by a factor around fifty.  This shows that our setup produces a strong amplification of the Drude-plasma discrepancy, which makes one confident  that an unambiguous discrimination between the two models should be possible with this apparatus.  
The  recent  isoelectronic differential experiment at IUPUI \cite{decca7}, using a  Au coated sphere  of radius $R \simeq 150 \mu$m, achieved a sensitivity better than 0.3 fN in force differences,  independent of the separation $a$ in the range from 200  nm to 1 $\mu$m.   With  this level of  sensitivity, it would be possible to accurately measure $\Delta F$   up to separations of several $\mu$m. 
 \begin{figure}
\includegraphics{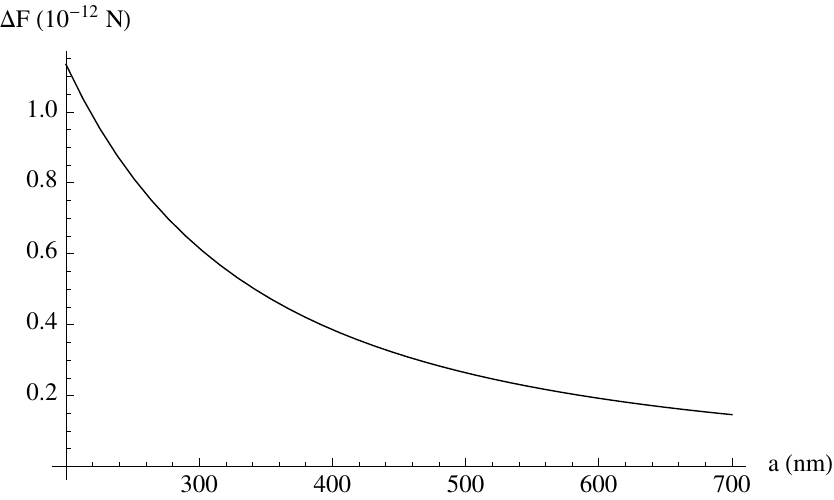}
\caption{\label{fig4}     Force difference $\Delta F$ (pN) versus separation $a$ (in nm) for a Ni-coated sphere and a Au-Ni plate, covered by 80 nm of Au.  The curve was computed using the Drude model with $\mu_{\rm Ni}(0)=110$.}
\end{figure}
\begin{figure}
\includegraphics{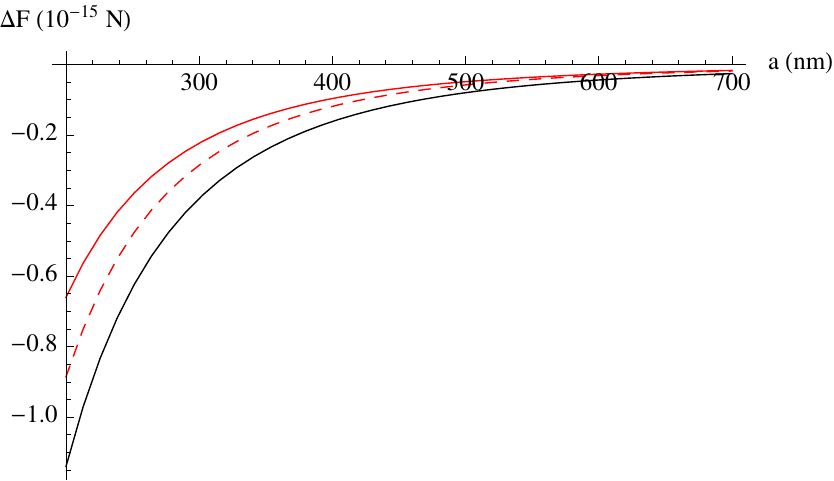}
\caption{\label{fig5}     Force difference $\Delta F$ (in fN) versus separation $a$ (in nm) for a Ni-coated sphere and a Au-Ni plate, covered by 80 nm of Au. The  black line  corresponds to the Drude model with $\mu_{\rm Ni}(0)=1$, while the solid and dashed red lines correspond to the plasma model, with $\mu_{\rm Ni}(0)=110$ and  $\mu_{\rm Ni}(0)=1$, respectively.}
\end{figure} 
\begin{figure}
\includegraphics{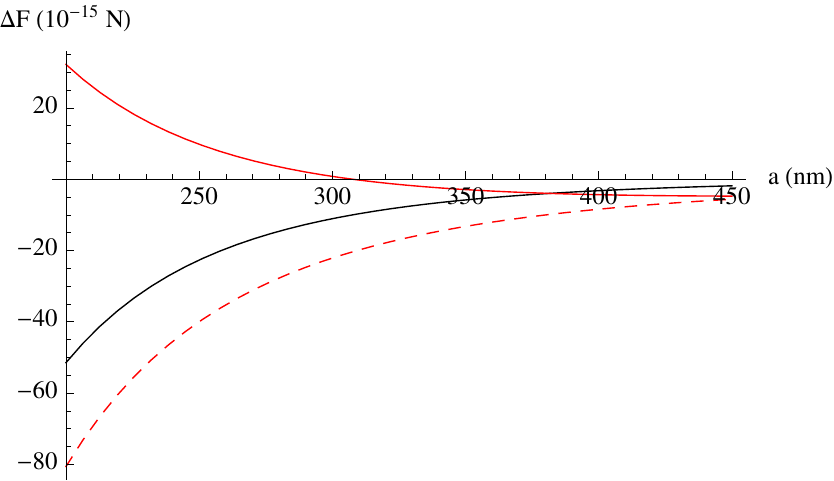}
\caption{\label{fig6bis}     Force difference $\Delta F$ (in fN) versus separation $a$ (in nm) for a Ni-coated sphere and a Pt-Ni plate, covered with 20 nm of Pt. The  black line  corresponds to the Drude model with $\mu_{\rm Ni}(0)=1$, while the solid and dashed red lines correspond to the plasma model, with $\mu_{\rm Ni}(0)=110$ and  $\mu_{\rm Ni}(0)=1$, respectively.}
\end{figure}
\begin{figure}
\includegraphics{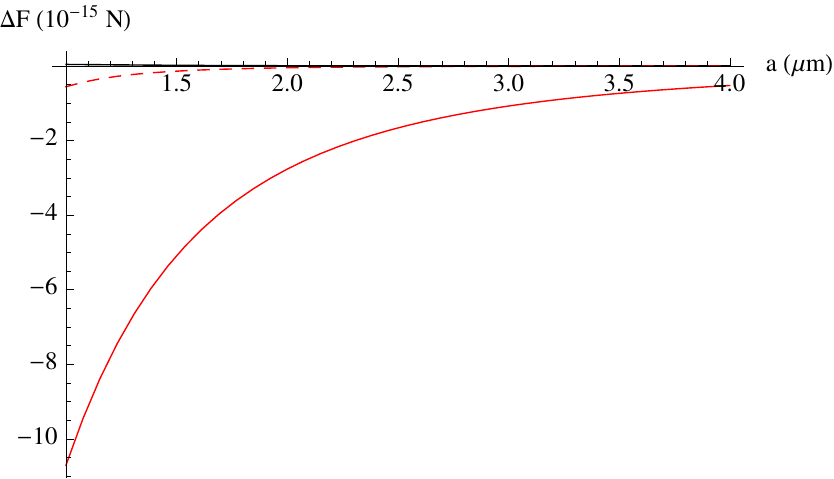}
\caption{\label{fig6}     Force difference $\Delta F$ (in fN) versus separation $a$ (in microns) for a Ni-coated sphere and a Pt-Ni plate, covered with 100 nm of conductive Si. The  black line  corresponds to the Drude model with $\mu_{\rm Ni}(0)=1$, while the solid and dashed red lines correspond to the plasma model, with $\mu_{\rm Ni}(0)=110$ and  $\mu_{\rm Ni}(0)=1$, respectively.}
\end{figure}
\begin{figure}
\includegraphics{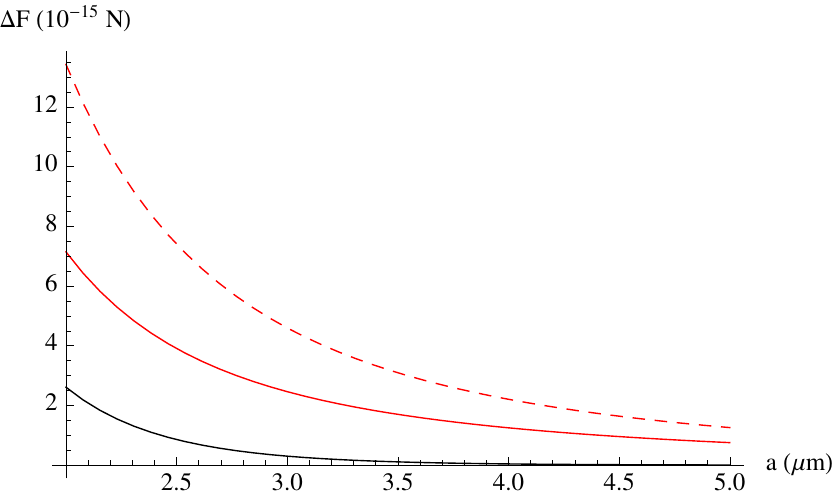}
\caption{\label{fig7}  Force difference $\Delta F$ (in fN) versus separation $a$ (in microns) for a Ni-coated sphere and a Si-Ni plate, covered with 100 nm of conductive Si. The  black line  corresponds to the Drude model with $\mu_{\rm Ni}(0)=1$, while the solid and dashed lines correspond to the plasma model, with $\mu_{\rm Ni}(0)=110$ and  $\mu_{\rm Ni}(0)=1$, respectively.  }
\end{figure}

\subsection{Magnetic metals}

In this Section we consider setups involving a ferromagnetic material. Magnetic materials bring up an interesting problem in connection with the Casimir effect: does the large  magnetic permeability of ferromagnetic substances  affect the Casimir force? To address this question,  we recall that the dynamic  magnetic permittivity of magnetic materials can be modeled by the  Debye formula \cite{vonso}
\be
\mu(\omega)=1+\frac{\mu(0)-1}{1-{\rm i}\, \omega/\omega_m}\;,
\ee 
where $\mu(0)$ is the static magnetic permeability and $\omega_m$ is a characteristic frequency, typically smaller than $10^5$ Hz. Since   $\omega_m$ is several orders of magnitude smaller than the frequency of the first Matsubara mode $\xi_1 \sim 10^{14}$  Hz at room temperature, when computing the Casimir free energy one can  set $\mu=1$ in all $l>0$ Matsubara terms in Eq. (\ref{lifs}).  This means that magnetic properties can affect the Casimir force only via the  $l=0$ TE term in Eq. (\ref{lifs}). The latter term is proportional to the temperature, and therefore it is clear that this term must describe the effect of  thermal fluctuations of the magnetic dipoles inside the ferromagnetic plates. However, it has been known for a long time that magnetic domains existing within (non magnetized) magnetic materials show thermal fluctuations only in small particles or at temperatures close to the Curie point $T_c$, while under normal circumstances the equilibrium magnetic microstructure is {\it athermal} \cite{hubert}. This makes one wonder whether magnetic properties affect the Casimir force at all, far from the Curie point. Below we shall consider specifically the case of Ni. This ferromagnetic material has a static permeability $\mu_{\rm Ni}(0)=110$, and a Curie temperature  $T_c=620$ K.  In order to address the question of the influence of the static permeability of Ni on the Casimir force,  we shall consider within both the Drude and the plasma prescriptions two further alternatives to estimate the $l=0$ TE mode, i.e. to take $\mu_{\rm Ni}(0)=110$   or instead to take $\mu_{\rm Ni}(0)=1$, the two alternatives corresponding respectively  to full inclusion or full neglect of the magnetic properties in the Casimir force. 

Recent experiments by the Riverside group  measured the gradient of the Casimir between a Au sphere and a Ni plate \cite{bani1}, and  between a Ni coated sphere and a Ni plate \cite{bani2}, in the separation range from 200 nm to 550 nm. The data of these experiments were interpreted as providing evidence for the influence of the magnetic properties of Ni on the Casimir force, and to be in agreement with the plasma prescription but  not with the Drude model. As it is usual in  Casimir experiments operating in the submicron range, the relative difference between the plasma and the Drude models  was found to be rather small, less than about ten percent. Therefore, the theoretical interpretation of the data required a delicate analysis of several possible sources of error. 

We show in this Section that   isoelectronic setups   strongly enhance the discrepancy between the Drude and the plasma models in magnetic systems, both for submicron separations and for micron separations of the plates.  We shall consider several such setups, with the purpose of obtaining a complete discrimination between the four different possible theoretical models that were considered in \cite{bani1,bani2}, i.e. the Drude and the plasma models with account or neglect of the magnetic properties of Ni.  

The first setup is designed to discriminate the Drude model with $\mu_{\rm Ni}(0)=110$ from the three other models: it uses Ni both for the sphere coating and the region C of the plate, while the overlayer and region B are both made of Au. The permittivity  $\epsilon_{\rm Au}({\rm i} \xi)$  of Au was computed by the same procedure described in the previous Section.  The permittivity $\epsilon_{\rm Ni}({\rm i} \xi)$  of Ni was computed by a similar procedure, using the tabulated optical data of \cite{palik}  extrapolated to low frequencies via a Drude model with parameters $\omega_{\rm Ni}=4.89\, {\rm eV}/\hbar$ and $\gamma_{\rm Ni}=0.0436 \, {\rm eV}/\hbar$ \cite{bani2}.  In Figs. \ref{fig4} and  \ref{fig5}  we plot $\Delta F$  versus separation $a$ (in nm), for a sphere of radius $R=150\; \mu$m and for a thickness $w=80$ nm of the Au overlayer. The curve in Fig. \ref{fig4} was computed using the Drude prescription with $\mu_{\rm Ni}(0)=110$, while the three curves in Fig. \ref{fig5} were computed using the Drude prescription with $\mu_{\rm Ni}(0)=1$ (black line), and the plasma prescription with $\mu_{\rm Ni}(0)=110$ (red line) and $\mu_{\rm Ni}(0)=1$ (dashed red line). Remarkably, we see that the Drude model with $\mu_{\rm Ni}(0)=110$ predicts a signal whose magnitude is three orders of magnitude larger than that predicted by the three other models.  Therefore, by this setup is  should be easy to discriminate the Drude model with with  $\mu_{\rm Ni}(0)=110$ from the other three models.  An experiment based on this scheme is presently ongoing  at IUPUI. The setup is identical to that described in \cite{decca7}, apart from the replacement of the gold coated sphere by  a Ni coated one, and the replacement in the rotating disk of the Si sectors by Ni sectors. Preliminary data presented by R.S. Decca at a conference in Canc\'un (Mexico) show that for a sample using a 84 nm Au overlayer $\Delta F$ is of the order of -1 fN (with an error of 0.3. fN) in the range of separations   from 200 nm to 400 nm. The observed $\Delta F$ has the opposite sign and is off by three orders of magnitudes with respect to the prediction of the Drude model with  $\mu_{\rm Ni}(0)=110$, while it is in qualitative agreement with the remaining three models. This rules out conclusively the Drude model with full account of the Ni permeability. It remains to see if the Ni permeability has any impact at all on the Casimir force.  To achieve a good discrimination between the remaining three models that we consider, it is necessary to change some of the materials used in the first setup. We consider first a setup in which Au   is replaced by  Pt, both for the overlayer and for region B. The permittivity $\epsilon_{\rm Pt}({\rm i} \xi)$  of Pt was computed by the same procedures described earlier, using the tabulated optical data of \cite{palik}  extrapolated to low frequencies via a Drude model with parameters $\omega_{\rm Pt}=5.1\, {\rm eV}/\hbar$ and $\gamma_{\rm Pt}=0.07 \, {\rm eV}/\hbar$ \cite{ordal}. In Fig. \ref{fig6bis} we show a plot of $\Delta F$ (in FN) versus separation (in nm) for a sphere of radius $R=150\; \mu$m, and for a thickness $w=20$ nm of the Pt overlayer. We see that in the range from 200 to 300 nm the plasma model with $\mu_{\rm Ni}(0)=110$ predicts a positive $\Delta F$, while both the Drude and the plasma models with $\mu_{\rm Ni}(0)=1$ predict a negative $\Delta F$. Such a sign difference should be easy to detect. Observation of a positive $\Delta F$ would provide a clear signature of the influence of the magnetic permeability on the Casimir force.   Another setup that could allow for a clear discrimination of the plasma model with $\mu_{\rm Ni}(0)=110$ versus the two other models with $\mu_{\rm Ni}(0)=1$ uses  a 100 nm overlayer made of conductive  Si, and again a region B  made of Pt.   In Fig. \ref{fig6} we show a plot of $\Delta F$ (in FN) versus separation (in micron) for a sphere of radius $R=150\; \mu$m. We see from the plot that with this setup also it should be easily possible to discriminate the  $\mu_{\rm Ni}(0)=110$ plasma model (solid red line) from the two $\mu_{\rm Ni}(0)=1$  models (black and dashed red line). A good discrimination among the latter two models is finally achieved by the third setup, which uses a Ni coated sphere in front of a plate made of dielectric Si (B region) and Ni (C region), covered by a conductive Si overlayer.  In Fig. \ref{fig7} we show a plot of $\Delta F$ (in FN) versus separation (in micron) for a sphere of radius $R=150\; \mu$m and for a thickness $w=100$ nm of the ${\rm Si}_c$ overlayer. We see that with this setup there is indeed a large difference between the signals predicted by  $\mu_{\rm Ni}(0)=1$  Drude model (grey curve) and the
$\mu_{\rm Ni}(0)=1$  plasma model  (dashed red-line). We conclude that the combined use of the three Ni setups should allow to determine whether or not the magnetic permeability of ferromagnetic substances   has any infuence on the Casimir effect, and  which among the plasma and Drude models better describes it.

\section{Conclusions}

The problem of thermal corrections in the Casimir effect  still constitutes a major unresolved puzzle in this field. Different theoretical prescriptions have been proposed to compute the thermal Casimir force, which provide different predictions for the thermal force. Several experiments have been performed in recent years to clarify the matter, but the situation remains unclear. Some experiments \cite{bani2,decca4,decca5,decca6,chang2} 
probing the Casimir force for separations between 200 and 700 nm have been interpreted as supporting the plasma prescription, and to rule out the Drude prescription, while a single experiment \cite{lamorth} probing the wide range from 700 nm to 7.3 micron was interpreted to be consistent with the Drude model, and inconsistent with the plasma model. A major problem in the theoretical interpretation of Casimir experiments is represented by stray electrostatic forces caused by potential patches on the plates. This represents a severe problem, especially for micron separations of the plates where stray  forces become large. 

An experimental approach which is in principle immune from the problem of patch potentials is based on isoelectronic force difference measurements. This approach has been successfully used by the IUPUI group in recent searches of non-newtonian gravitational forces in the submicron range \cite{deccaiso,decca7}. The author showed recently \cite{hide,hide2} that the isoelectronic technique can be adapted to investigate the thermal Casimir effect. In this paper we have further developed this idea. We have shown that by suitably choosing the materials of the samples, it is possible to investigate in great detail the features of the thermal Casimir force, both for magnetic and non-magnetic metals. Apart from being immune from electrostatic problems, the devices described in this paper provide a strong amplification of the differences between the Drude and the plasma prescriptions, thus making a discrimination among these models much simpler  and hopefully conclusive.      
 
Preliminary results of an ongoing experiment at IUPUI with Ni samples \cite{experdecca}, based on the isoelectronic scheme described in this paper,  already prove conclusively that the effect of the magnetic permeability    on  the Casimir force  is three orders of magnitude smaller than what can be predicted on the basis of the  Drude model. The available  data are  qualitatively consistent with the plasma model. Another possibility to interpret the data is that magnetic propertites do not affect at all the Casimir effect, in which case both the Drude and the plasma models are qualitatively in agreement with the data.        In this paper we have described three different isolectronic setups that should allow to fully resolve this problem, and to establish which among the Drude and the plasma models provides the correct description of the Casimir effect in magnetic systems.

\section{Acknowledgments}

The author thanks R.S. Decca for valuable discussions while the manuscript was in preparation.


\section*{References}

 



\begin{thebibliography}{200}

\bibitem{Casimir48} H.~B.~G. Casimir, Proc. K. Ned. Akad. Wet. {\bf 51}, 793 (1948).

\bibitem{lifs} E. M. Lifshitz, Zh. Eksp. Teor. Fiz. {\bf 29}, 94 (1955) [Sov. Phys. JETP {\bf 2}, 73 (1956)].

\bibitem{rytov} S.M.\ Rytov, {\it Theory of Electrical Fluctuations and Thermal Radiation}, Publyshing House, Academy os Sciences, USSR (1953).

\bibitem{book1} K. A. Milton, {\it The Casimir Effect: Physical manifestations of Zero-Point Energy} (World Scientific, Singapore, 2001).

\bibitem{parse} V. A. Parsegian, {\it Van der Waals Forces} (Cambridge University Press, Cambridge, England,
2005).


\bibitem{book2} M. Bordag, G. L. Klimchitskaya, U. Mohideen, and V. M. Mostepanenko,  {\it Advances in the Casimir Effect} (Oxford University Press, Oxford, 2009).

\bibitem{revexp} G. L. Klimchitskaya, U. Mohideen, and V. M. Mostepanenko,  Rev. Mod. Phys.{\bf 81},  1827 (2009).

\bibitem{lamor1} S.K. Lamoreaux, Phys. Rev. Lett. {\bf 78}, 5 (1997).

\bibitem{umar}  U. Mohideen and A. Roy, Phys. Rev. Lett. {\bf 81}, 4549 (1998).

\bibitem{gianni} G. Bressi, G. Carugno, R. Onofrio and G. Ruoso, Phys. Rev. Lett. {\bf 88}, 041804 (2002).

\bibitem{decca4} R. S. Decca, D. L\'{o}pez, E. Fischbach et al, Annals Phys. {\bf 318}, 37 (2005).

\bibitem{decca5} R. S. Decca, D.  L\'{o}pez, E. Fischbach et al, Phys. Rev. D {\bf 75}, 077101 (2007).

\bibitem{decca6} R. S. Decca, D.  L\'{o}pez, E. Fischbach et al, Eur. Phys. J. C {\bf 51},  963 (2007).

\bibitem{chang2} C.-C. Chang, A.A. Banishev, R. Castillo-Garza et al, Phys. Rev. B {\bf 85}, 165443 (2012).

\bibitem{semic} F. Chen, U.  Mohideen, G.L. Klimchitskaya, and V.M. Mostepanenko,  Phys. Rev. A {\bf 72}, 020101(R) (2005); ibid.  {\bf 74}, 022103 (2006).

\bibitem{ito} S. de Man, K. Heeck, R. J. Wijngaarden, and D. Iannuzzi, Phys. Rev. Lett. {\bf 103}, 040402  (2009).

\bibitem{bani1} A. A. Banishev, C.-C. Chang, G. L. Klimchitskaya, V. M. Mostepanenko, and U. Mohideen, Phys. Rev. B {\bf 85}, 195422 (2012).

\bibitem{bani2} A. A. Banishev, G. L. Klimchitskaya, V. M. Mostepanenko, and U. Mohideen, Phys. Rev. Lett.  {\bf 110}, 137401 (2013);  Phys. Rev. B {\bf 88}, 155410 (2013).

\bibitem{liq} J. N. Munday, F. Capasso and V. A. Parsegian,  Nature {\bf 457}, 170 (2009).

\bibitem{ala1}   G. Bimonte, E. Calloni, G. Esposito, L. Milano, and L. Rosa,  Phys. Rev. Lett. {\bf 94}, 180402  (2005).

\bibitem{ala2} G. Bimonte, E. Calloni, G. Esposito, and L. Rosa, Nucl. Phys. B {\bf 726}, 441 (2005).

\bibitem{super1} G. Bimonte, Phys. Rev.  A {\bf 78}, 062101  (2008).

\bibitem{superc} G. Bimonte, D. Born, E. Calloni, G. Esposito, U. Huebner, E. Il'Ichev, L. Rosa, F. Tafuri, and R. Vaglio, J. Phys. A {\bf 41}, 164023 (2008).

\bibitem{capasso} H.B. Chan, V.A. Aksyuk, R.N. Kleiman, D.J. Bishop, and F. Capasso, Science {\bf 291}, 1941 (2001). 

\bibitem{chen} F. Chen,  U. Mohideen, G. L. Klimchitskaya, and V. M. Mostepanenko Phys. Rev. Lett., {\bf 88}, 101801 (2002).

\bibitem{chan} H. B. Chan, Y. Bao, J. Zou, R. A. Cirelli, F. Klemens, W. M. Mansfield, and C. S. Pai, Phys. Rev. Lett. {\bf 101},  030401 (2008).

\bibitem{bao} Y. Bao, R. Gu\'{e}rout, J. Lussange, A. Lambrecht, R. A. Cirelli, F. Klemens, W. M. Mansfield, C. S. Pai, and H. B. Chan Phys. Rev. Lett. {\bf 105}, 250402 (2010).

\bibitem{chiu} H. -C. Chiu, G. L. Klimchitskaya, V. N. Marachevsky, V. M. Mostepanenko, and U. Mohideen, Phys. Rev. B {\bf 80}, 121402(R) (2009);  Phys. Rev. B {\bf 81}, 115417 (2010).


\bibitem{bani} A. A. Banishev, J. Wagner, T. Emig, R. Zandi and U. Mohideen, Phys. Rev. Lett. {\bf 110}, 250403 (2013).

\bibitem{deccanat}  F. Intravaia, S. Koev,  I.W. Jung, A.A. Talin P.S. Davids, R.S. Decca,  V.A. Aksyuk and  D.A.R.  Dalvit, and D.  L\'{o}pez,  Nat. Comm.  {\bf 4},  2515 (2013).


 

 

\bibitem{capassorev} A. W. Rodriguez, P.-C. Hui, D. P. Woolf, S.G. Johnson, M. Lon\u{c}ar,  and F. Capasso, Ann. Phys. (Berlin) {\bf 527}, 45 (2015).

\bibitem{bezerra} V. B. Bezerra, G. L. Klimchitskaya and V. M. Mostepanenko, Phys. Rev. A {\bf 65}, 052113 (2002); ibid. {\bf 66}, 062112 (2002).

\bibitem{bezerra2} V.B. Bezerra, G.L. Klimchitskaya, V.M. Mostepanenko and C. Romero, Phys. Rev. A {\bf 69}, 022119 (2004).

\bibitem{Martin} P. R. Buenzli and Ph. A. Martin, Phys. Rev. E {\bf 77}, 011114 (2008). 

\bibitem{bimo2} G. Bimonte, Phys. Rev. A {\bf 79}, 042107 (2009).

 \bibitem{critiz} G. L. Klimchitskaya, M. Bordag, and V. M. Mostepanenko, Int. J. Mod. Phys. A {\bf 27},  1260012 (2012).

\bibitem{antonini} P. Antoninii, G. Bimonte, G. Bressi, G. Carugno, G. Galeazzi, G. Messineo, and  G. Ruoso, J. Phys. Conf. Ser. {\bf 161}, 012006 (2009).

\bibitem{lamorth}  A. O. Sushkov, W. J. Kim, D. A. R. Dalvit, and S. K. Lamoreaux, Nature Phys. {\bf 7}, 230 (2011).


\bibitem{deccapatch} R.O. Behunin, D.A.R. Dalvit, R.S. Decca, C. Genet, I. W. Jung, A. Lambrecht, A. Liscio, D. L\'opez, S. Reynaud, G. Schnoering, 
G. Voisin, and Y. Zeng,  Phys. Rev. A {\bf 90}, 062115 (2014).
 

 
\bibitem{hide} G. Bimonte, Phys. Rev. Lett. {\bf 112}, 240401 (2014).

\bibitem{hide2} G. Bimonte, Phys. Rev. Lett. {\bf 113}, 240405 (2014).


\bibitem{deccaiso} R. S. Decca,  D. L\'{o}pez, H. B. Chan, E. Fischbach, D. E. Krause, and C. R. Jamell, Phys. Rev. Lett. {\bf 94}, 240401 (2005).

\bibitem{decca7} Y. J.-Chen, W.K. Tham, D.E. Krause,  D. L\'{o}pez, E. Fischbach, and R.S. Decca, arXiv:1410.7267. 


\bibitem{speake} R. O. Behunin, D.A.R. Dalvit, R.S. Decca, and C.C Speake, Phys. Rev. D {\bf 89}, 051301(R) (2014).

\bibitem{experdecca} R.S. Decca,  slides of the talk at the Les Houches conference {\it Casimir Physics} (Les Houches 2014)  http://www.spectro.jussieu.fr/Casimir-Physics-Talks.
 
\bibitem{Piro} I. Pirozhenko, A. Lambrecht, and V. B. Svetovoy, New. J. Phys. {\bf 8}, 238 (2006).

\bibitem{sveto} V. B. Svetovoy, P. J. van Zwol, G. Palasantzas, and J. Th. M. De Hosson, Plys. Rev. B {\bf 77}, 035439 (2008).

\bibitem{generKK} G. Bimonte, Phys. Rev A {\bf 81}, 062501 (2010).

\bibitem{generKK2} G. Bimonte, Phys. Rev A  {\bf 83}, 042109 (2011).

\bibitem{super2} G. Bimonte, H. Haakh, C. Henkel and F. Intravaia, J. Phys. A {\bf 43}, 145304 (2010).

\bibitem{john} G. Bimonte, New. J. Phys. {\bf 9} 281 (2007).

\bibitem{carsten} F. Intravaia and C. Henkel, Phys. Rev. Lett. {\bf 103}, 130405 (2009).



\bibitem{kruger} M. Kr\"uger, V. A. Golyk, G. Bimonte, and M. Kardar, EPL {\bf 104}, 41001 (2013). 

\bibitem{fosco2}  C. D. Fosco, F. C. Lombardo, and F. D. Mazzitelli, Phys. Rev.D {\bf 84}, 105031 (2011).

\bibitem{bimonte3} G. Bimonte, T. Emig, R. L. Jaffe, and M. Kardar, EPL {\bf 97}, 50001 (2012).

\bibitem{bimonte4} G. Bimonte, T. Emig, and M. Kardar, Appl. Phys. Lett. {\bf 100}, 074110 (2012).

\bibitem{bimonteCP} G. Bimonte, T. Emig, and M. Kardar, Phys. Rev. D {\bf 90}, 081702(R) (2014).

\bibitem{richmond} P. Richmond and B. W. Ninham, J. Phys. C {\bf 4}, 1988 (1971).

\bibitem{tomas} M. S. Toma\u{s}, Phys. Lett. A {\bf 342}, 381 (2005).


\bibitem{palik}  {\it Handbook of Optical Constants of Solids}, edited by E. D. Palik (Academic, New York, 1995).

\bibitem{vonso} S. V. Vonsovskii, {\it Magnetism} (Wiley, New York, 1974).

\bibitem{hubert} A. Hubert and R. Sch\"afer, {\it Magnetic Domains} (Springer, 1998).

\bibitem{ordal} M.A. Ordal, R.J. Bell, R.W. Alexander, L.L. Long, and M. R. Querry, App. Opt. {\bf 24}, 4493 (1985).










\end{thebibliography}
\end{document}